\pdfoutput=1
\documentclass[conference]{IEEEtran}
\IEEEoverridecommandlockouts
\usepackage{cite}
\usepackage{amsmath,amssymb,amsfonts}

\newtheorem{theorem}{Theorem}
\newtheorem{lemma}{Lemma}
\newtheorem{proposition}{Proposition}

\usepackage{algorithmic}
\usepackage{graphicx}
\usepackage{textcomp}
\usepackage{xcolor}
\def\BibTeX{{\rm B\kern-.05em{\sc i\kern-.025em b}\kern-.08em
    T\kern-.1667em\lower.7ex\hbox{E}\kern-.125emX}}
\begin{document}

\title{ORBGRAND: Achievable Rate for General Bit Channels and Application in BICM\\
\thanks{This work was supported in part by the National Natural Science Foundation of China under Grant 62231022.}
}

\author{\IEEEauthorblockN{Zhuang Li and Wenyi Zhang}
\IEEEauthorblockA{\text{Department of Electronic Engineering and Information Science} \\
\text{University of Science and Technology of China}\\
Hefei, China \\
Email: lizhuang06@mail.ustc.edu.cn, wenyizha@ustc.edu.cn}

}

\maketitle

\begin{abstract}
Guessing random additive noise decoding (GRAND) has received widespread attention recently, and among its variants, ordered reliability bits GRAND (ORBGRAND) is particularly attractive due to its efficient utilization of soft information and its amenability to hardware implementation. It has been recently shown that ORBGRAND is almost capacity-achieving in additive white Gaussian noise channels under antipodal input. In this work, we first extend the analysis of ORBGRAND achievable rate to memoryless binary-input bit channels with general output conditional probability distributions. The analytical result also sheds insight into understanding the gap between the ORBGRAND achievable rate and the channel mutual information. As an application of the analysis, we study the ORBGRAND achievable rate of bit-interleaved coded modulation (BICM). Numerical results indicate that for BICM, the gap between the ORBGRAND achievable rate and the channel mutual information is typically small, and hence suggest the feasibility of ORBGRAND for channels with high-order coded modulation schemes.
\end{abstract}


\section{Introduction}
Guessing random additive noise decoding (GRAND) \cite{duffy2019capacity} \cite{riaz2022universal} has been recently proposed as a universal decoding paradigm. Its basic idea is to sort and test a sequence of possible error patterns until finding a valid codeword. If all error patterns are sorted from the most likely to the least likely, then GRAND is, in fact, equivalent to maximum-likelihood (ML) decoding.\par
Utilizing soft symbol reliability information can improve decoding performance \cite[Ch. 10]{Lin2004ErrorCC}. Symbol reliability GRAND (SRGRAND) \cite{duffy2021guessing} uses one bit of soft information to specify
whether a channel output symbol is reliable. Soft GRAND (SGRAND) \cite{solomon2020soft} uses magnitudes of channel output to generate error patterns, so it can make full usage of channel soft information and it in fact implements the maximum-likelihood decoding. But SGRAND requires a sequential online algorithm to generate the sequence of error patterns, and this limits its efficiency of hardware implementation. Another variant of GRAND, called ordered reliability bits GRAND (ORBGRAND) \cite{duffy2022ordered}, does not require exact values of channel output, and instead uses only the relationship of ranking among channel outputs to generate error patterns. This property enables ORBGRAND to generate error patterns offline and facilitates hardware implementation \cite{condo2021high} \cite{abbas2022high} \cite{condo2022fixed}. Noteworthily, it has been shown via an information theoretic analysis in \cite{liu2022orbgrand} that, ORBGRAND achieves an information rate almost approaching the channel mutual information in additive white Gaussian noise (AWGN) channels under antipodal input.\par
Several variants of GRAND have been proposed for fading channels. In \cite{sarieddeen2022grand}, ORBGRAND based on reduced-complexity pseudo-soft information has been studied. In \cite{abbas2022grand}, fading-GRAND has been proposed for Rayleigh fading channels, shown to outperform traditional hard-decision decoders, and in \cite{abbas2023hardware}, a hardware architecture of fading-GRAND has been studied. In \cite{chatzigeorgiou2022symbol}, symbol-level GRAND has been proposed for block fading channels, utilizing the knowledge of modulation scheme and channel state information (CSI). In \cite{allahkaram2022urllc} \cite{allahkaram2023symbol}, GRAND has been applied to multiple-input-multiple-output channels.\par 
In this paper, along the line of \cite{liu2022orbgrand}, we conduct an information theoretic analysis to study the performance limit of ORBGRAND, for general memoryless binary-input bit channels. The motivation is that, achievable rates for general bit channels serve as basic building blocks for assessing the performance of ORBGRAND for channels with high-order coded modulation schemes, such as bit-interleaved coded modulation (BICM) \cite{caire1998bit} \cite{i2008bit}, a technique extensively used in fading channels. Since ORBGRAND is a mismatched decoding method, rather than the maximum-likelihood one, we utilize generalized mutual information (GMI) to quantify its achievable rate \cite{ganti2000mismatched} \cite{lapidoth2002fading}. A comparison between the derived GMI of ORBGRAND and the channel mutual information also sheds insight into understanding the gap between them, helping explain when and why the GMI of ORBGRAND is close to the channel mutual information, as usually observed in numerical studies. As an application of the analysis, we study the ORBGRAND achievable rate of BICM. Numerical results for QPSK, 8PSK, and 16QAM with Gray and set-partitioning labelings over AWGN and Rayleigh fading channels are presented. These results indicate that for BICM, the gap between the ORBGRAND achievable rate and the channel mutual information is typically small, and hence suggest the feasibility of ORBGRAND for channels with high-order coded modulation schemes.\par
The remaining part of this paper is organized as follows: Section \ref{system model} introduces the system model and ORBGRAND for general memoryless binary-input bit channels. Section \ref{achievable rate} derives the GMI of ORBGRAND and discusses the gap between it and the channel mutual information. Section \ref{simulation results} presents the corresponding numerical results for BICM. Section \ref{conclusion} concludes this paper. 
\section{System Model and ORBGRAND for General Bit Channels}
\label{system model}
\subsection{System Model}
\label{channel model}
In this paper, we study memoryless binary-input channels with general output conditional probability distributions. Without loss of generality, let the input alphabet be $\{+1, -1\}$, and let the output probability distribution be $q_+(y)$ under input $x = +1$ and $q_-(y)$ under input $x = -1$, respectively. Note that $q_+(y)$ and $q_-(y)$ are general and we do not require them to possess any symmetric property.\par
We consider a codebook with code length $N$ and code rate $R$ nats per channel use, so the number of messages is $M =\lceil e^{NR}\rceil$. When sending message $m$, the transmitted codeword is $\underline{x}(m) = [x_1(m),x_2(m),\cdots,x_N(m)]$. We assume that the elements of $\underline{x}(m)$ are indepedent and identically distributed (i.i.d.) uniform $\{+1, -1\}$ random variables. This is a common assumption in random coding analysis, and is satisfied for many linear codes. The channel output vector is denoted as $\underline{\mathsf{Y}} = [\mathsf{Y}_1,\mathsf{Y}_2,\cdots,\mathsf{Y}_N]$. Define the log-likelihood ratio (LLR) random variable $\mathsf{T}_n=\text{ln}\frac{q_{+}(\mathsf{Y}_n)}{q_{-}(\mathsf{Y}_n)}$ and the reliability vector $[\lvert \mathsf{T}_1\rvert,\lvert \mathsf{T}_2\rvert,\cdots,\lvert \mathsf{T}_N\rvert]$. For $n = 1,2,\cdots,N$, denote $\mathsf{R}_n$  as the rank of $\lvert\mathsf{T}_n\rvert$ among the sorted array consisting of $\left\{\lvert \mathsf{T}_1\rvert,\lvert \mathsf{T}_2\rvert,\cdots,\lvert \mathsf{T}_N\rvert\right\} $, from $1$ (the smallest) to $N$ (the largest). The cumulative distribution function (cdf) of $\lvert \mathsf{T}\rvert$ is denoted as $\Psi(t)$, for $t \geq 0$. We use lowercase letters to denote the realizations of random variables; for example, $r_n$ as the realization of $\mathsf{R}_n$, and so on.
\subsection{ORBGRAND for General Bit Channels}
For a general bit channel, the procedure of ORBGRAND can be described as follows: when a channel output vector $\underline{y}$ is received, we calculate its reliability vector $\left[\lvert t_1\rvert, \lvert t_2\rvert, \cdots, \lvert t_N\rvert\right]$ and its hard-decision vector $\underline{x}_{\text{hard}}=[\text{sgn}(t_1), \text{sgn}(t_2), \cdots, \text{sgn}(t_N)]$, where $t_i=\text{ln}\frac{q_{+}(y_i)}{q_{-}(y_i)}$ is the LLR and $\text{sgn}(t) = 1$ if $t \geq 0$ and $-1$ otherwise. We generate an error pattern matrix $P$ of size $2^N\times N$. The elements of $P$ are $+1$ or $-1$, and the rows of $P$ are all distinct: if the $q$-th row $n$-th column element $P_{q, n}$ is $-1$, then in the $q$-th query, we flip the sign of the $n$-th element of $\underline{x}_\mathrm{hard}$. So each row of $P$ represents a different query and we conduct the queries from top to bottom, until finding a valid codeword and declaring it as the decoded codeword, or exhausting all the rows without finding a valid codeword and declaring a decoding failure. We arrange the rows of $P$ so that the sum reliability of the $q$-th row defined as $\sum_{n: P_{q, n} = -1} r_n$ is non-decreasing with $q$, where $r_n$ is the rank of $|t_n|$ introduced in the previous subsection. There exist efficient algorithms for generating the matrix $P$; see, e.g., \cite{duffy2022ordered} \cite{condo2021high}. Since $2^N$ is typically an exceedingly large quantity, in practice we can truncate the matrix to keep only its first $Q$ rows, where $Q$ is the maximum number of queries permitted.\par
As shown in \cite{liu2022orbgrand},  the following form of decoding criterion provides a unified description of GRAND and its variants including ORBGRAND, if $Q$ is set to its maximum possible value $2^N$: for a received channel output vector $\underline{y}$, the decoder decides the message to be
 \begin{equation}
 	\begin{aligned}
 	\label{decoding rule}
 	\hat{m} &= \mathop{\arg\min}\limits_{m = 1, 2,\cdots, M}\frac{1}{N}\sum_{n=1}^{N}\\
 	&\quad\quad\quad\gamma_n(\underline{y})\cdot \mathbf{1}\left(\text{sgn}\left(\text{ln}\frac{q_{+}(y_n)}{q_{-}(y_n)}\right)\cdot x_n(m)\textless 0\right).
 	\end{aligned}
 \end{equation}
It can be shown (for details see, e.g., \cite[Sec. II]{liu2022orbgrand}) that the decoding criterion (\ref{decoding rule}) produces the same decoding result as, and is hence equivalent to, GRAND and its variants, if we are permitted to conduct an exhaustive query of all possible error patterns, i.e., $Q = 2^N$. \footnote{As mentioned in the previous paragraph, in practice $Q$ is usually set as a number smaller than $2^N$, but the form of (\ref{decoding rule}) renders an information theoretic analysis amenable, as will be seen in the next section.} Different choices of $\left\{\gamma_n\right\}_{n=1,2,\cdots,N}$ correspond to different sorting criteria of error patterns: if $\gamma_n(\underline{y})=1$, (\ref{decoding rule}) is the original GRAND \cite{duffy2019capacity};
if $\gamma_n(\underline{y})=\left\lvert \text{ln}\frac{q_{+}(y_n)}{q_{-}(y_n)}\right\rvert$, (\ref{decoding rule}) is SGRAND \cite{solomon2020soft}, which is equivalent to the maximum-likelihood decoding;
if $\gamma_n(\underline{y})=\frac{r_n}{N}$, where $r_n$ is the realization of the rank random variable $\mathsf{R}_n$, (\ref{decoding rule}) is ORBGRAND \cite{duffy2022ordered}. \par
\section{GMI Analysis of ORBGRAND}
\label{achievable rate}
\subsection{GMI of ORBGRAND}
 Since ORBGRAND is not the maximum likelihood decoding, we resort to mismatched decoding analysis (see, e.g., \cite{ganti2000mismatched} \cite{lapidoth2002fading}) for characterizing its information theoretic performance limit. For this purpose, GMI is a convenient tool and has been widely used. GMI quantifies the maximum rate such that the ensemble average probability of decoding error asymptotically vanishes as the code length grows without bound. For general memoryless bit channels, the GMI of ORBGRAND is characterized by the following theorem.
\begin{theorem}
	\label{theorem}
	For the system setup in Section \ref{system model}, the GMI of ORBGRAND is given by
	\begin{equation}
		\begin{aligned}
			I_{\text{ORBGRAND}} &= \text{ln}2 - \inf\limits_{\theta \textless 0}\Bigg\{\int_{0}^{1} \text{ln}(1 + e^{\theta t})dt\\
			&\quad -\theta\cdot\frac{1}{2}\int_{q_{+}(y)\textless q_{-}(y)}\Psi\left(\left\lvert \text{ln}\frac{q_{+}(y)}{q_{-}(y)}\right\rvert\right) q_{+}(y)dy\\
			&\quad -\theta\cdot\frac{1}{2}\int_{q_{+}(y)\textgreater q_{-}(y)}\Psi\left(\left\lvert \text{ln}\frac{q_{+}(y)}{q_{-}(y)}\right\rvert\right)q_{-}(y)dy\Bigg\}
			\label{ORBGRAND rate}
		\end{aligned}    
	\end{equation}
	in nats/channel use.\par
\end{theorem}
\begin{IEEEproof}
Since in ORBGRAND, the terms inside the summation in (\ref{decoding rule}) are correlated due to the relationship of ranking, we cannot directly invoke the standard formula of GMI (see, e.g., \cite[Eqn. (12)]{ganti2000mismatched}) to evaluate the GMI of ORBGRAND. Instead, we conduct analysis and calculation from the first principle, similar to \cite{liu2022orbgrand} which considers the special case of AWGN channels only. We calculate the ensemble average probability of decoding error. As a consequence of i.i.d. random coding, the average probability of decoding error is equal to the probability of decoding error under the condition of transmitting message $m=1$.\par
Based on the general decoding rule (\ref{decoding rule}), we define the decoding metric of ORBGRAND by
	\begin{equation}
		\begin{aligned}
				\mathsf{D}(m)&=\frac{1}{N}\sum_{n=1}^{N}\\
				&\quad\frac{\mathsf{R}_n}{N}\cdot \mathbf{1}\left(\text{sgn}\left(\mathsf{T}_n\right)\cdot \mathsf{X}_n(m)\textless 0\right), \quad m=1, 2, \cdots, M.
				\label{decoding metric}
		\end{aligned}
	\end{equation}
Under i.i.d random coding, $\left\{\lvert \mathsf{T}_n\rvert\right\}_{n=1,2,\cdots,N}$ are also i.i.d. with cdf $\Psi(t)$, for $t \geq 0$.\par
When the transmitted message is $m=1$, we can characterize the asymptotic behavior of the decoding metric in (\ref{decoding metric}) using the following three lemmas, whose proofs are placed in Appendix \ref{A}.
\begin{lemma}
	\label{lemma1}
	As $N \rightarrow \infty$, for the transmitted codeword, the expectation of the decoding metric in (\ref{decoding metric}) is given by
	\begin{equation}
		\begin{aligned}
			\lim\limits_{N\to+\infty} \mathbb{E} \mathsf{D}(1) &= \frac{1}{2}\int_{q_{+}(y)\textless q_{-}(y)}\Psi\left(\left\lvert \text{ln}\frac{q_{+}(y)}{q_{-}(y)}\right\rvert\right) q_{+}(y)dy\\
			&\quad +\frac{1}{2}\int_{q_{+}(y)\textgreater q_{-}(y)}\Psi\left(\left\lvert \text{ln}\frac{q_{+}(y)}{q_{-}(y)}\right\rvert\right)q_{-}(y)dy.  
		\end{aligned}     
	\end{equation}	
\end{lemma}
\begin{lemma}
	\label{lemma2}
	As $N \rightarrow \infty$, for the transmitted codeword, the variance of the decoding metric in (\ref{decoding metric}) is given by
	\begin{equation}
		\lim\limits_{N\to+\infty} \text{var} \mathsf{D}(1) =  0.          
	\end{equation}
\end{lemma}
\begin{lemma}
	\label{lemma3}
	As $N \rightarrow \infty$, for any codeword not transmitted, i.e., $m^{\prime}\neq 1$ and any $\theta  \textless 0$, the decoding metric in (\ref{decoding metric}) satisfies, almost surely,
	\begin{equation}
		\begin{aligned}
			\Delta(\theta) & := \lim\limits_{N\to+\infty} \frac{1}{N}\text{ln}\mathbb{E}\left\{e^{N \theta \mathsf{D}(m^{\prime})}\Big\lvert \underline{\mathsf{T}}\right\}\\
			& = \int_{0}^{1} \text{ln}(1 + e^{\theta t})dt - \text{ln}2.
		\end{aligned}
	\end{equation}
\end{lemma}\par
For any $\epsilon \textgreater 0$, we define event\par
\begin{equation*}
	\begin{aligned}
		\mathcal{U}_{\epsilon} &= \Bigg\{\mathsf{D}(1) \geq \frac{1}{2}\int_{q_{+}(y)\textless q_{-}(y)}\Psi\left(\left\lvert \text{ln}\frac{q_{+}(y)}{q_{-}(y)}\right\rvert\right) q_{+}(y)dy\\
		&\quad +\frac{1}{2}\int_{q_{+}(y)\textgreater q_{-}(y)}\Psi\left(\left\lvert \text{ln}\frac{q_{+}(y)}{q_{-}(y)}\right\rvert\right)q_{-}(y)dy + \epsilon \Bigg\},
	\end{aligned}
\end{equation*}
so the ensemble average probability of decoding error is
\begin{equation}
	\begin{aligned}
		& \text{Pr}[\hat{m} \neq 1]\\
		=\quad & \text{Pr}[\hat{m} \neq 1\lvert \mathcal{U}_\epsilon]\text{Pr}[\mathcal{U}_\epsilon] + \text{Pr}[\hat{m} \neq 1\lvert \mathcal{U}_\epsilon^c]\text{Pr}[\mathcal{U}_\epsilon^c] \\
		\leq \quad & \text{Pr}[\mathcal{U}_\epsilon] + \text{Pr}[\hat{m} \neq 1\lvert \mathcal{U}_\epsilon^c].
		\label{Pr}
	\end{aligned}
\end{equation}\par
Using Lemma \ref{lemma1}, Lemma \ref{lemma2} and Chebyshev's inequality, we can deduce that for any $\epsilon \textgreater 0$,\par
\begin{small}
\begin{equation}
	\begin{aligned}
		\lim\limits_{N\to+\infty}&\text{Pr}\Bigg[\mathsf{D}(1) \geq \frac{1}{2}\int_{q_{+}(y)\textless q_{-}(y)}\Psi\left(\left\lvert \text{ln}\frac{q_{+}(y)}{q_{-}(y)}\right\rvert\right) q_{+}(y)dy\\
		&\quad +\frac{1}{2}\int_{q_{+}(y)\textgreater q_{-}(y)}\Psi\left(\left\lvert \text{ln}\frac{q_{+}(y)}{q_{-}(y)}\right\rvert\right)q_{-}(y)dy + \epsilon\Bigg]=0;
	\end{aligned}
\end{equation}
\end{small}%
this shows that $\text{Pr}[\mathcal{U}_\epsilon]$ can be arbitrarily close to zero as the code length $N$ grows without bound.\par
Meanwhile, based on the decoding rule (\ref{decoding rule}) and the union bound, we have\par
\begin{small}
\begin{equation}
	\begin{aligned}
		& \text{Pr}[\hat{m} \neq 1\lvert \mathcal{U}_\epsilon^c]\\
		\leq \quad &\text{Pr}\Bigg[\exists m^{\prime}\neq 1, \mathsf{D}(m^{\prime})\textless\frac{1}{2}\int_{q_{+}(y)\textless q_{-}(y)}\Psi\left(\left\lvert \text{ln}\frac{q_{+}(y)}{q_{-}(y)}\right\rvert\right) q_{+}(y)dy\\
		&\quad +\frac{1}{2}\int_{q_{+}(y)\textgreater q_{-}(y)}\Psi\left(\left\lvert \text{ln}\frac{q_{+}(y)}{q_{-}(y)}\right\rvert\right)q_{-}(y)dy + \epsilon\Bigg]\\
		\leq \quad & e^{NR} \text{Pr}\Bigg[\mathsf{D}(m^{\prime}) \textless \frac{1}{2}\int_{q_{+}(y)\textless q_{-}(y)}\Psi\left(\left\lvert \text{ln}\frac{q_{+}(y)}{q_{-}(y)}\right\rvert\right) q_{+}(y)dy\\
		&\quad +\frac{1}{2}\int_{q_{+}(y)\textgreater q_{-}(y)}\Psi\left(\left\lvert \text{ln}\frac{q_{+}(y)}{q_{-}(y)}\right\rvert\right)q_{-}(y)dy + \epsilon\Bigg].
		\label{PrUc}
	\end{aligned}
\end{equation}
\end{small}\par
Considering the conditional version of the probability in (\ref{PrUc}), and applying Chernoff's bound, we have that for any $N$ and any $\theta\textless 0$,\par
\begin{small}
\begin{equation}
	\begin{aligned}
		& -\frac{1}{N}\text{ln} \text{Pr}\Bigg[\mathsf{D}(m^{\prime}) \textless \frac{1}{2}\int_{q_{+}(y)\textless q_{-}(y)}\Psi\left(\left\lvert \text{ln}\frac{q_{+}(y)}{q_{-}(y)}\right\rvert\right) q_{+}(y)dy\\
		&\quad +\frac{1}{2}\int_{q_{+}(y)\textgreater q_{-}(y)}\Psi\left(\left\lvert \text{ln}\frac{q_{+}(y)}{q_{-}(y)}\right\rvert\right)q_{-}(y)dy + \epsilon\Big\lvert \underline{\mathsf{T}} \Bigg]\\
		\quad \geq \quad &\theta\Bigg[\frac{1}{2}\int_{q_{+}(y)\textless q_{-}(y)}\Psi\left(\left\lvert \text{ln}\frac{q_{+}(y)}{q_{-}(y)}\right\rvert\right) q_{+}(y)dy+\frac{1}{2}\int_{q_{+}(y)\textgreater q_{-}(y)}\\
		& \quad\Psi\left(\left\lvert \text{ln}\frac{q_{+}(y)}{q_{-}(y)}\right\rvert\right)q_{-}(y)dy + \epsilon\Bigg] -\frac{1}{N}\text{ln}\mathbb{E}\left\{e^{N \theta \mathsf{D}(m^{\prime})}\Big\lvert \underline{\mathsf{T}}\right\}.
	\end{aligned}
\end{equation}
\end{small}\par Letting $\epsilon\rightarrow 0$, the code length $N\rightarrow +\infty$, and applying the almost surely limit in Lemma \ref{lemma3}, we have\par
\begin{small}
\begin{equation}
	\begin{aligned}
		&\text{Pr}\Bigg[\mathsf{D}(m^{\prime}) \textless \frac{1}{2}\int_{q_{+}(y)\textless q_{-}(y)}\Psi\left(\left\lvert \text{ln}\frac{q_{+}(y)}{q_{-}(y)}\right\rvert\right) q_{+}(y)dy\\
		&\quad +\frac{1}{2}\int_{q_{+}(y)\textgreater q_{-}(y)}\Psi\left(\left\lvert \text{ln}\frac{q_{+}(y)}{q_{-}(y)}\right\rvert\right)q_{-}(y)dy + \epsilon\Big\lvert \underline{\mathsf{T}} \Bigg]\\
		\quad \leq \quad&\exp\Bigg\{-N\Bigg[\theta\cdot \frac{1}{2}\int_{q_{+}(y)\textless q_{-}(y)}\Psi\left(\left\lvert \text{ln}\frac{q_{+}(y)}{q_{-}(y)}\right\rvert\right) q_{+}(y)dy\\
		&\quad +\theta\cdot\frac{1}{2}\int_{q_{+}(y)\textgreater q_{-}(y)}\Psi\left(\left\lvert \text{ln}\frac{q_{+}(y)}{q_{-}(y)}\right\rvert\right)q_{-}(y)dy\\
		&\quad -\int_{0}^{1} \text{ln}(1 + e^{\theta t})dt + \text{ln}2\Bigg]\Bigg\}.
		\label{Pr1}
	\end{aligned}
\end{equation}
\end{small}\par Substituting (\ref{Pr1}) into (\ref{PrUc}), and applying the law of total expectation to remove the conditioning, we assert that the ensemble average probability of decoding error asymptotically vanishes as the code length $N$ grows without bound if the code rate $R$ satisfies the following inequality, i.e., (\ref{ORBGRAND rate}) in the statement of Theorem \ref{theorem}.\par
\begin{small}
\begin{equation}
	\begin{aligned}
		R \quad\textless\quad & \sup\limits_{\theta \textless 0}\Bigg\{\theta\cdot\frac{1}{2}\int_{q_{+}(y)\textless q_{-}(y)}\Psi\left(\left\lvert \text{ln}\frac{q_{+}(y)}{q_{-}(y)}\right\rvert\right) q_{+}(y)dy\\
		&\quad +\theta\cdot\frac{1}{2}\int_{q_{+}(y)\textgreater q_{-}(y)}\Psi\left(\left\lvert \text{ln}\frac{q_{+}(y)}{q_{-}(y)}\right\rvert\right)q_{-}(y)dy\\
		&\quad -\int_{0}^{1} \text{ln}(1 + e^{\theta t})dt + \text{ln}2 \Bigg\}\\
		\quad=\quad& \text{ln}2 - \inf\limits_{\theta \textless 0}\Bigg\{\int_{0}^{1} \text{ln}(1 + e^{\theta t})dt\\
		&\quad -\theta\cdot\frac{1}{2}\int_{q_{+}(y)\textless q_{-}(y)}\Psi\left(\left\lvert \text{ln}\frac{q_{+}(y)}{q_{-}(y)}\right\rvert\right) q_{+}(y)dy\\
		&\quad -\theta\cdot\frac{1}{2}\int_{q_{+}(y)\textgreater q_{-}(y)}\Psi\left(\left\lvert \text{ln}\frac{q_{+}(y)}{q_{-}(y)}\right\rvert\right)q_{-}(y)dy\Bigg\}.
	\end{aligned}    
\end{equation}
\end{small}
\end{IEEEproof}

\subsection{Discussion on Gap between GMI of ORBGRAND and Channel Mutual Information}
\label{discussion}
For general memoryless bit channels under uniform binary input, as described in Section \ref{channel model}, the channel mutual information is given by 
\begin{equation}
	\begin{aligned}
		I&= I(\mathsf{X};\mathsf{Y})\\
		&=\text{ln}2-\frac{1}{2}\int_{-\infty}^{+\infty}\text{ln}\left(1+\frac{q_{-}(y)}{q_{+}(y)}\right)q_{+}(y)dy\\
		&\quad -\frac{1}{2}\int_{-\infty}^{+\infty}\text{ln}\left(1+\frac{q_{+}(y)}{q_{-}(y)}\right)q_{-}(y)dy
		\label{channel capacity}
	\end{aligned}
\end{equation}
in nats/channel use.\par
In order to analyze the gap between (\ref{ORBGRAND rate}) and (\ref{channel capacity}), we rewrite (\ref{channel capacity}) into its equivalent form as the GMI of SGRAND, noticing that SGRAND is equivalent to the maximum-likelihood decoding and thus achieves the channel mutual information. This leads to the following result.\par
\begin{proposition}
	For the general memoryless bit channel model in Section \ref{system model}, the GMI of SGRAND is given by
	\begin{equation}
		\begin{aligned}
			I_{\text{SGRAND}}&=\text{ln}2-\inf\limits_{\theta \textless0}\Bigg\{\mathbb{E}\left[\text{ln}\left(1+e^{\theta \left\lvert\text{ln}\frac{q_{+}(\mathsf{Y})}{q_{-}(\mathsf{Y})}\right\rvert}\right)\right]\\
			&\quad -\theta \cdot \frac{1}{2}\int_{q_{+}(y)\textless q_{-}(y)}\left\lvert\text{ln}\frac{q_{+}(y)}{q_{-}(y)}\right\rvert q_{+}(y)dy\\
			&\quad -\theta \cdot\frac{1}{2}\int_{q_{+}(y)\textgreater q_{-}(y)}\left\lvert\text{ln}\frac{q_{+}(y)}{q_{-}(y)}\right\rvert q_{-}(y)dy \Bigg\}
			\label{SGRAND rate}
		\end{aligned}
	\end{equation}
	in nats/channel use, and this is equal to the channel mutual information $I$ in (\ref{channel capacity}). 
\end{proposition}
\begin{IEEEproof}
	See Appendix \ref{B}.
\end{IEEEproof}

It is interesting to note that, if $\left\lvert\text{ln}\frac{q_{+}(\mathsf{Y})}{q_{-}(\mathsf{Y})}\right\rvert$ and $\left\lvert\text{ln}\frac{q_{+}(y)}{q_{-}(y)}\right\rvert$ in (\ref{SGRAND rate}) are replaced by $\Psi\left(\left\lvert\text{ln}\frac{q_{+}(\mathsf{Y})}{q_{-}(\mathsf{Y})}\right\rvert\right)$ and $\Psi\left(\left\lvert \text{ln}\frac{q_{+}(y)}{q_{-}(y)}\right\rvert\right)$ respectively, then (\ref{SGRAND rate}) will become (\ref{ORBGRAND rate}). This is obtained by noting that  $\Psi\left(\left\lvert\text{ln}\frac{q_{+}(\mathsf{Y})}{q_{-}(\mathsf{Y})}\right\rvert\right)$ obeys the uniform distribution over $[0, 1]$ and hence
\begin{equation}
	\mathbb{E}\left[\text{ln}\left(1+e^{\theta \Psi\left(\left\lvert\text{ln}\frac{q_{+}(\mathsf{Y})}{q_{-}(\mathsf{Y})}\right\rvert\right)}\right)\right]=\int_{0}^{1} \text{ln}(1 + e^{\theta t})dt.
\end{equation} 
Therefore, the gap between $I_{\text{ORBGRAND}}$ and $I$ is essentially caused by the difference between $t$ and $\Psi(t)$. If $\Psi(t)$ behaves close to a linear function, then the GMI of ORBGRAND will be close to the channel mutual information. \par
\begin{figure}[htbp]	\centerline{\includegraphics[width=0.38\textwidth]{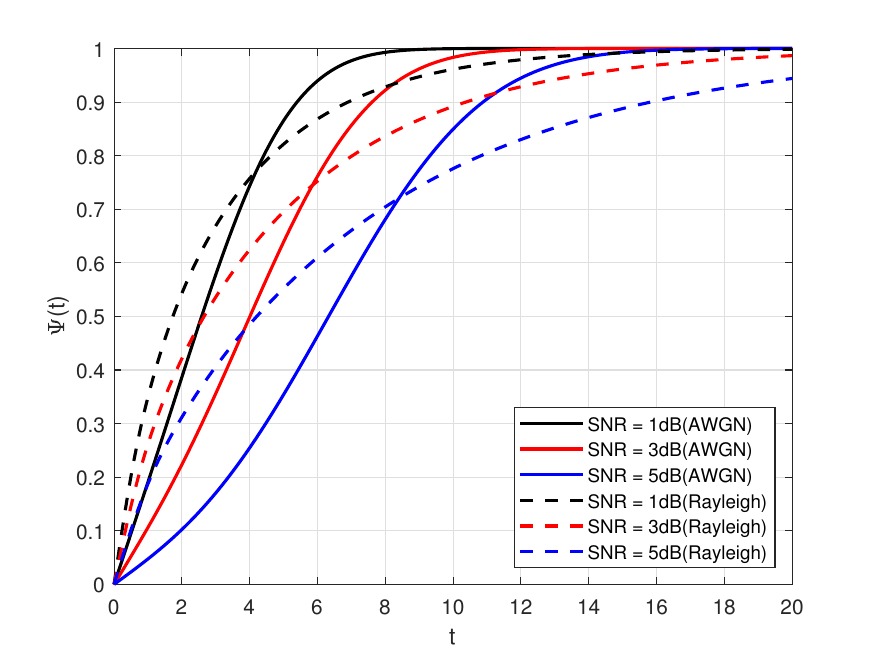}}
	\caption{Plots of $\Psi(t)$ under under AWGN and Rayleigh fading channels.}
	\label{fig:Psi}
\end{figure}
\begin{figure}[htbp]	\centerline{\includegraphics[width=0.38\textwidth]{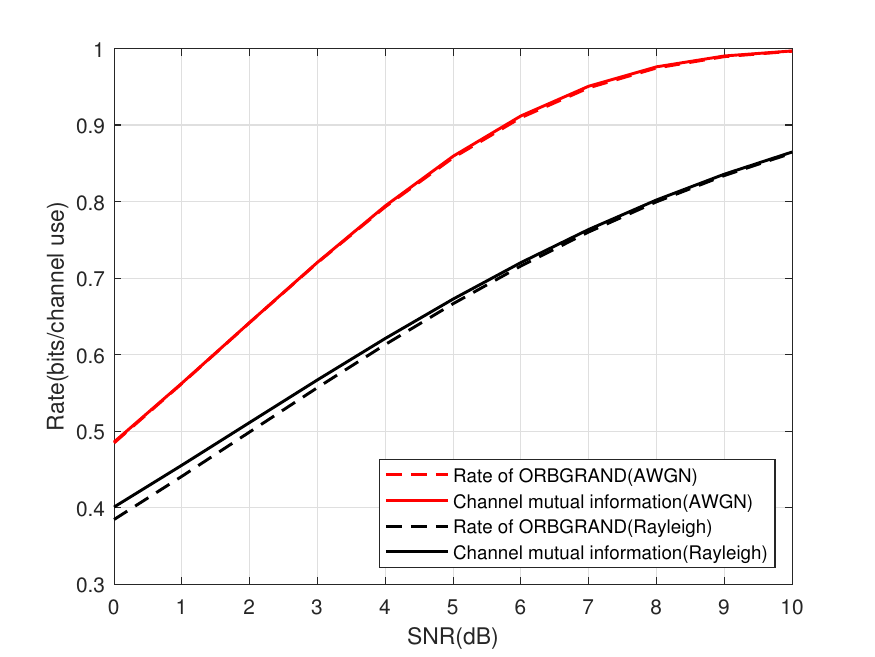}}
	\caption{Plots of $I_{\text{ORBGRAND}}$ and $I$ under AWGN and Rayleigh fading channels.}
	\label{fig:rate}
\end{figure}

Here we give a simple example to illustrate the above discussion. We consider BPSK modulation over the Rayleigh fading channel with perfect CSI and the AWGN channel. The curves of $\Psi(t)$ and $I_{\text{ORBGRAND}}$ under different values of signal-to-noise ratio (SNR) are displayed in Fig. \ref{fig:Psi} and Fig. \ref{fig:rate}, respectively. Taking $\text{SNR}=3\text{dB}$ as an example, we can see from Fig. \ref{fig:Psi} that the linearity of $\Psi(t)$ in the Rayleigh fading channel is obviously worse than that in the AWGN channel. Therefore, we can see from Fig. \ref{fig:rate} that there is a noticeable gap between $I_{\text{ORBGRAND}}$ and $I$ in the Rayleigh fading channel, while there is virtually no gap in the AWGN channel.
\section{Application in BICM}
\label{simulation results}
BICM is an effective coded modulation scheme and has been widely used in contemporary communication systems. In this section, we use the analytical results in the previous section to calculate the ORBGRAND achievable rate of BICM, and compare it with the channel mutual information. This study serves as a theoretical basis for the feasibility of ORBGRAND for channels with high-order coded modulation schemes. \par
\subsection{Experimental Setup}
\begin{figure}[htbp]	\centerline{\includegraphics[width=0.45\textwidth]{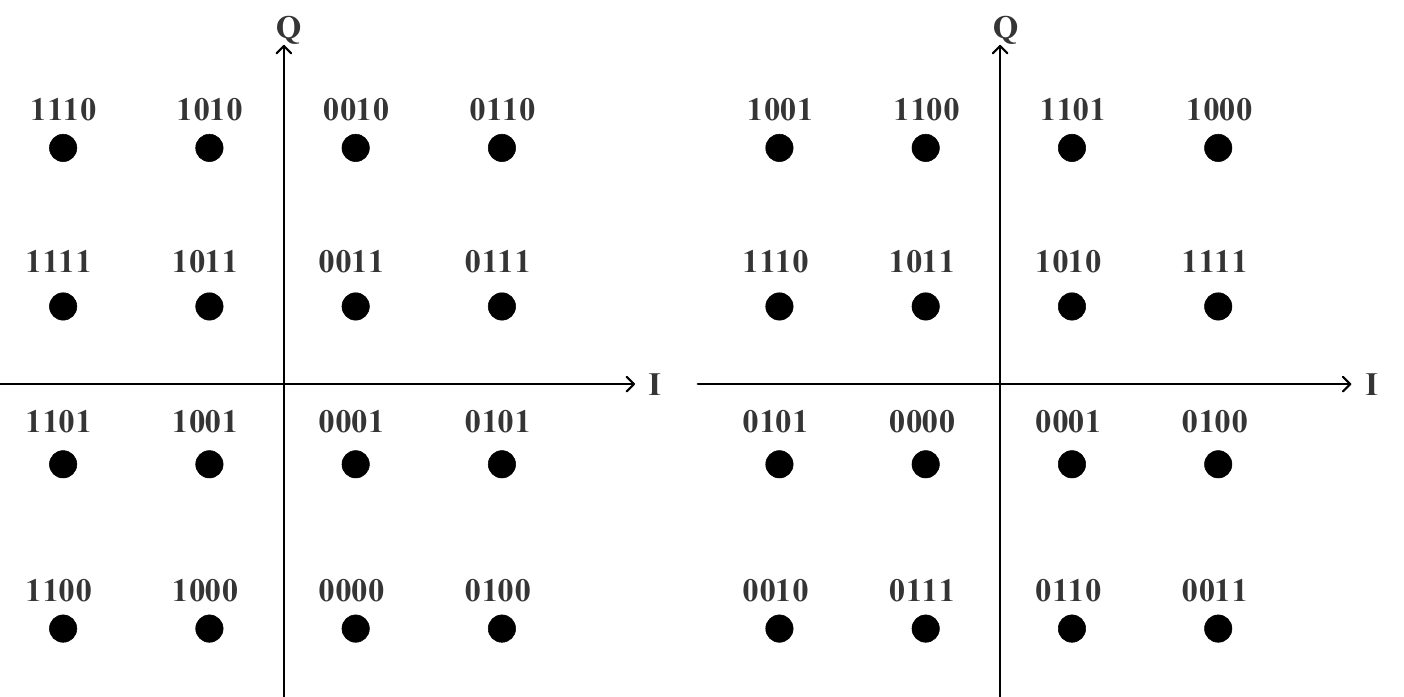}}
	\caption{Gray (left) and set-partitioning (right) labelings for 16QAM.}
	\label{fig:Gray and SP for 16QAM}
\end{figure}

In our experiment, we consider QPSK, 8PSK and 16QAM with ideal interleaving and perfect CSI. For each modulation type, we consider both Gray and set-partitioning labelings. For example, the constellation diagrams of the two labelings for 16QAM are shown in Fig. \ref{fig:Gray and SP for 16QAM}. The channel input-output relationship is
\begin{equation}
	\begin{aligned}
	\mathsf{Y} = \mathsf{H} \mathsf{S} + \mathsf{Z}.
	\label{BICM model}
	\end{aligned}
\end{equation}
$\mathsf{H}$ is the channel gain: when $\mathsf{H}=1$, (\ref{BICM model}) is the AWGN channel, and when $\mathsf{H}$ obeys a unit-variance circularly symmetric complex Gaussian distribution, (\ref{BICM model}) is the Rayleigh fading channel; $\mathsf{S}$ is the channel input, corresponding to a point in the constellation diagram; $\mathsf{Z}$ is the standard circularly symmetric complex Gaussian noise. In BICM, the codeword $\underline{\mathsf{X}}$ is first passed to an interleaver $\pi$, and the interleaved sequence $\pi(\underline{\mathsf{X}})$ is then divided into multiple subsequences, each of length matched to the order of the constellation, and is thus mapped to a point in the constellation diagram according to a certain labeling rule; for details, see, e.g., \cite{caire1998bit} \cite{i2008bit}.\par 
\begin{figure}[htbp]	\centerline{\includegraphics[width=0.31\textwidth]{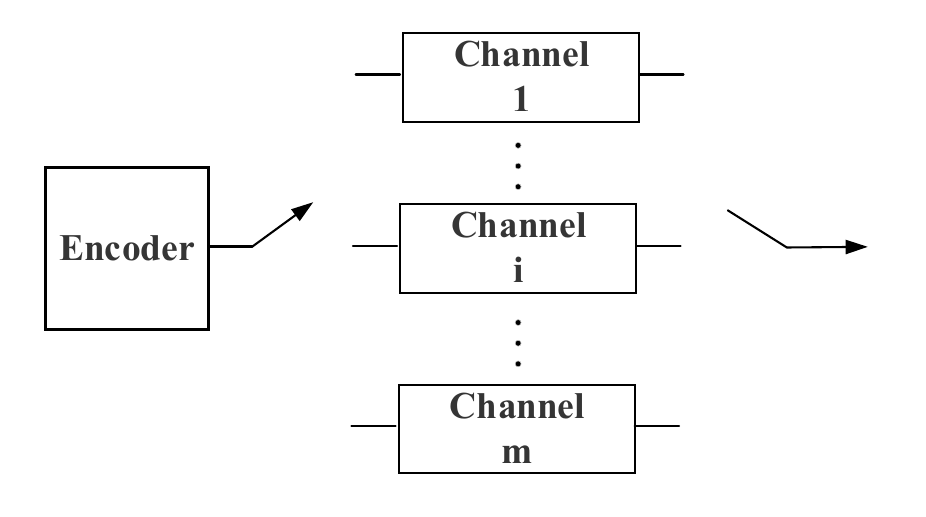}}
	\caption{Parallel channel model of BICM with ideal interleaving.}
	\label{fig:Parallel channel}
\end{figure}

Due to the nature of ideal interleaving, we can adopt the concept of parallel channel model in \cite{caire1998bit}, as shown in Fig. \ref{fig:Parallel channel}. The ORBGRAND achievable rate of the $i$-th parallel channel is denoted as $I^i_{\text{ORBGRAND}}$, so $I_{\text{ORBGRAND}}=\sum\limits_{i = 1}^mI^i_{\text{ORBGRAND}}$, where $m$ is the number of parallel channels. For the $i$-th parallel channel, the conditional probability distribution of $y$ is given by
\begin{gather}
		q^i_{+}(y)= \frac{\sum\limits_{s \in \mathcal{X}_1^i}p(y\lvert s)}{\lvert \mathcal{X}_1^i\rvert}, \quad q^i_{-}(y)= \frac{\sum\limits_{s \in \mathcal{X}_0^i}p(y\lvert s)}{\lvert \mathcal{X}_0^i\rvert},\notag \\
		\label{q}
\end{gather}
where $\mathcal{X}_1^i$ is the set of $s$ whose $i$-th bit is $1$, and $\mathcal{X}_0^i$ is the set of $s$ whose $i$-th bit is $0$. \footnote{Here, $1$ (resp. $0$) corresponds to $+1$ (resp. $-1$) in our bit channel model in Sections \ref{system model} and \ref{achievable rate}.} Plugging (\ref{BICM model}) and (\ref{q}) into (\ref{ORBGRAND rate}), we can calculate $I^i_{\text{ORBGRAND}}$, and thus get $I_{\text{ORBGRAND}}$.\par
\subsection{Numerical Results}
In general, the ORBGRAND achievable rate and the channel mutual information in BICM do not yield closed-form expressions, so we use numerical methods such as Monte Carlo to evaluate them. \par
\begin{figure}[htbp]	\centerline{\includegraphics[width=0.38\textwidth]{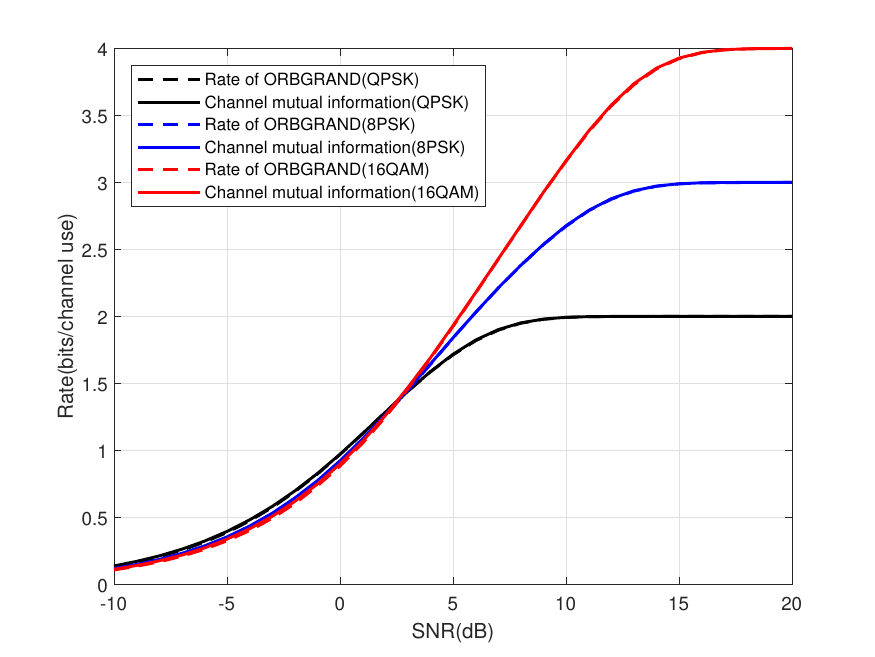}}
	\caption{ORBGRAND achievable rate and channel mutual information under QPSK, 8PSK and 16QAM for AWGN channel in the case of Gray labeling.}
	\label{fig:Gray_AWGN}
\end{figure}
\begin{figure}[htbp]	\centerline{\includegraphics[width=0.38\textwidth]{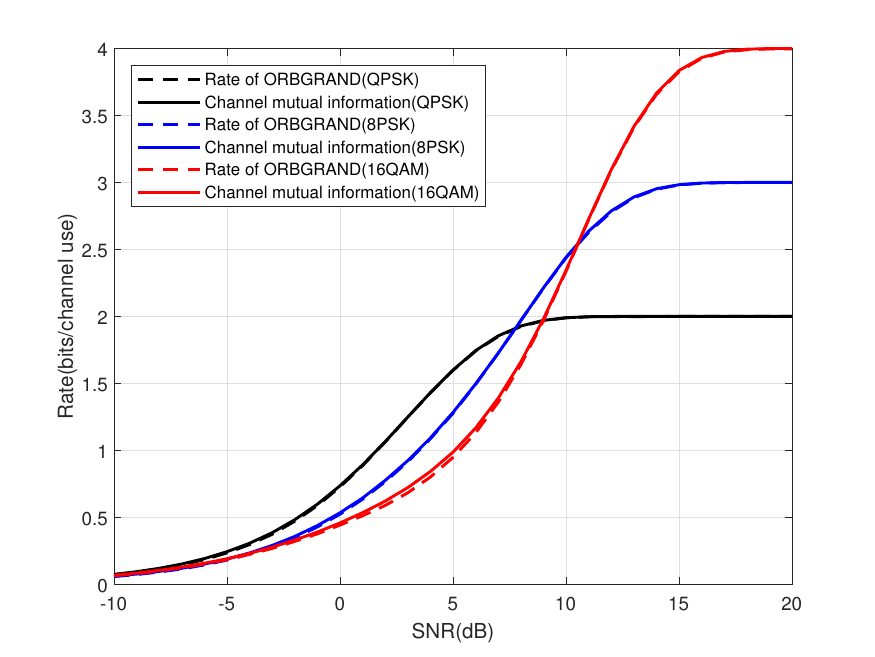}}
	\caption{ORBGRAND achievable rate and channel mutual information under QPSK, 8PSK and 16QAM for AWGN channel in the case of set-partitioning labeling.}
	\label{fig:SP_AWGN}
\end{figure}
\begin{figure}[htbp]	\centerline{\includegraphics[width=0.38\textwidth]{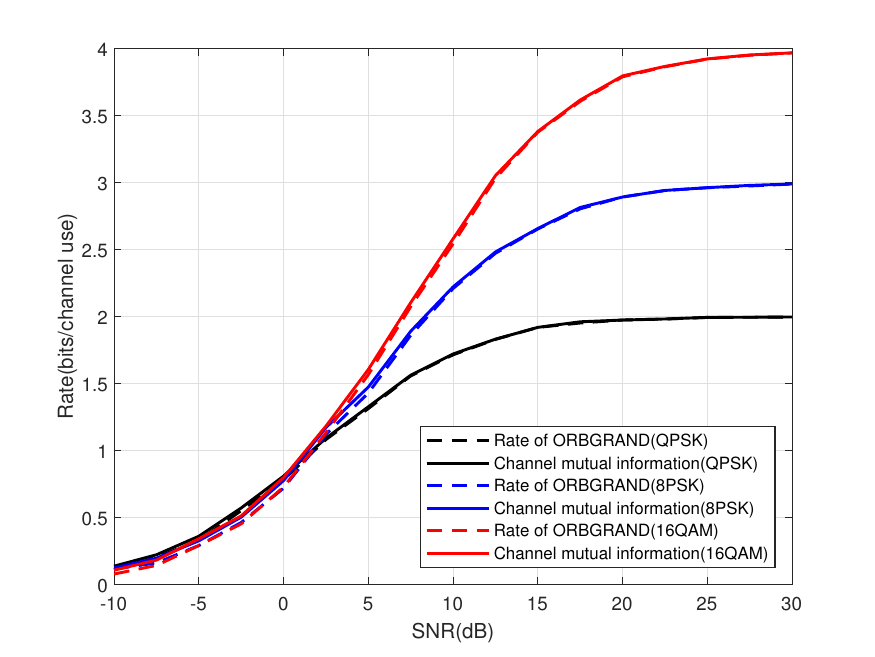}}
	\caption{ORBGRAND achievable rate and channel mutual information under QPSK, 8PSK and 16QAM for Rayleigh fading channel with perfect CSI in the case of Gray labeling.}
	\label{fig:Gray_Rayleigh}
\end{figure}
\begin{figure}[htbp]	\centerline{\includegraphics[width=0.38\textwidth]{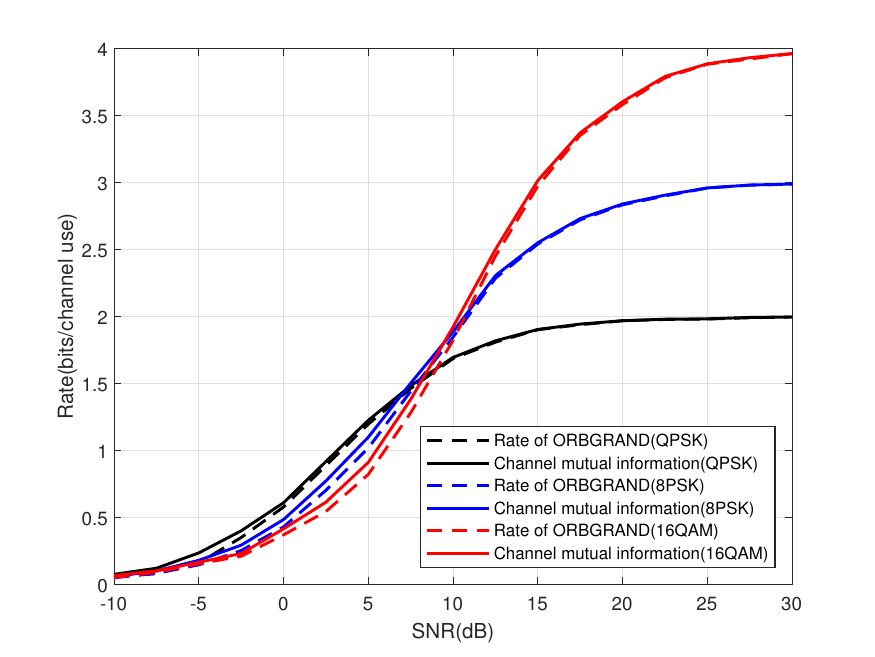}}
	\caption{ORBGRAND achievable rate and channel mutual information under QPSK, 8PSK and 16QAM for Rayleigh fading channel with perfect CSI in the case of set-partitioning labeling.}
	\label{fig:SP_Rayleigh}
\end{figure}

The numerical results for the AWGN channel are shown in Fig. \ref{fig:Gray_AWGN} and Fig. \ref{fig:SP_AWGN}, which show that although ORBGRAND is a mismatched decoder, the ORBGRAND achievable rate under QPSK, 8PSK and 16QAM over the AWGN channel is very close to the channel mutual information, regardless of the labeling. The numerical results for the Rayleigh fading channel are shown in Fig. \ref{fig:Gray_Rayleigh} and Fig. \ref{fig:SP_Rayleigh}, which exhibit essentially the same trend as that in the AWGN channel, with a slightly larger gap between the ORBGRAND achievable rate and the channel mutual information in the low SNR regime. As discussed in Section \ref{discussion}, the gap is due to the nonlinearity of the cdf of the magnitude of the channel LLR. These numerical results suggest that ORBGRAND can still maintain good decoding performance for channels adopting high-order coded modulation schemes.\par

\section{Conclusion}
\label{conclusion}
In this paper, we conduct an achievable rate analysis of ORBGRAND for memoryless binary-input channels with general output conditional probability distributions. The achievable rate is characterized by the GMI of ORBGRAND, and its analysis sheds insight into why and when the GMI is close to the channel mutual information, a phenomenon usually observed in several representative channels of practical interest. This analysis further paves the way towards analyzing the performance of ORBGRAND for high-order coded modulation schemes, such as BICM. Numerical results for BICM indicate the near-optimal performance of ORBGRAND, and thus suggest its feasibility for high-rate transmission systems, where high-order modulations are necessary. \par
\appendices\section{}\label{A}
\section*{Proof of lemmas 1-3}
\subsection{Proof of Lemma 1}
We have
\begin{equation}
	\begin{aligned}
		\label{ED1}
		&\mathbb{E}\mathsf{D}(1)=\frac{1}{N^2} \sum_{n = 1}^N \mathbb{E}[\mathsf{R}_n \mathbf{1}(\text{sgn}(\mathsf{T}_n)\cdot  \mathsf{X}_n(1)\textless 0)].   
	\end{aligned}
\end{equation}
For each term in (\ref{ED1}), we have 
\begin{equation}
	\begin{aligned}
		&\mathbb{E}[\mathsf{R}_n \mathbf{1}(\text{sgn}(\mathsf{T}_n)\cdot  \mathsf{X}_n(1)\textless 0)]\\
		\quad=\quad& \frac{1}{2}\mathbb{E}\Big[\mathsf{R}_n \mathbf{1}(\mathsf{T}_n \textless 0)\Big\lvert \mathsf{X}_n(1)=+1\Big]\\
		&+ \frac{1}{2}\mathbb{E}\Big[\mathsf{R}_n \mathbf{1}(\mathsf{T}_n \textgreater 0)\Big\lvert \mathsf{X}_n(1)=-1\Big].
		\label{term} 
	\end{aligned}
\end{equation}\par
For the first expectation in (\ref{term}), based on the law of total expectation, we have
\begin{equation}
	\begin{aligned}
		&\mathbb{E}[\mathsf{R}_n \mathbf{1}(\mathsf{T}_n \textless 0)\lvert \mathsf{X}_n(1)=+1]\\
		\quad=\quad&\mathbb{E}\Big[\mathbb{E}\Big[\mathsf{R}_n\mathbf{1}(q_{+}(\mathsf{Y}_n)\textless q_{-}(\mathsf{Y}_n)) \Big\lvert \mathsf{X}_n(1)=+1,\mathsf{Y}_n\Big]\Big].
	\end{aligned}
\end{equation}
Next we have 
\begin{align}
	\nonumber
	&\mathbb{E}\Big[\mathsf{R}_n\mathbf{1}(q_{+}(\mathsf{Y}_n)\textless q_{-}(\mathsf{Y}_n)) \Big\lvert \mathsf{X}_n(1)=+1,\mathsf{Y}_n=y\Big]\\
	 \quad=\quad& \left\{
	\begin{aligned}
		& \mathbb{E}\Big[\mathsf{R}_n \Big\lvert \mathsf{X}_n(1)=+1,\mathsf{Y}_n=y\Big] \quad \text{if}\quad q_{+}(y)\textless q_{-}(y),\\
		& 0 \quad\text{else}.\\
	\end{aligned}
	\right.
	\label{expectation}
\end{align}
Based on the definition of $\mathsf{R}_n$ and the i.i.d. nature of $\left\{\lvert \mathsf{T}_n\rvert\right\}_{n=1,2,\cdots,N}$, we notice that the expectation in the first branch of (\ref{expectation}) is exactly the expectaion of the rank when inserting $\left\lvert\text{ln}\frac{q_{+}(y)}{q_{-}(y)} \right\rvert$ into a sorted array of $N-1$ samples of $\left\{\lvert \mathsf{T}\rvert\right\}$. For simplicity, denoting the expectation in the first branch of (20) as $u(y)$, so
\begin{equation}
	\begin{aligned}
		&\mathbb{E}[\mathsf{R}_n \mathbf{1}(\mathsf{T}_n \textless 0)\lvert \mathsf{X}_n(1)=+1]=\int_{q_{+}(y)\textless q_{-}(y)}u(y) q_{+}(y)dy.
	\end{aligned}
\end{equation} \par
The second expectation in (\ref{term}) can be treated in the same approach. Therefore, we have
\begin{equation}
	\begin{aligned}
		\mathbb{E}\mathsf{D}(1)&= \frac{1}{2}\int_{q_{+}(y)\textless q_{-}(y)}\frac{u(y)} {N}q_{+}(y)dy\\
		&\quad+\frac{1}{2}\int_{q_{+}(y)\textgreater q_{-}(y)}\frac{u(y)} {N}q_{-}(y)dy.
	\end{aligned}
\end{equation}\par
Utilizing the asymptotic behavior of binomial distribution (for details see \cite[Appendix F]{liu2022orbgrand}), we obtain
\begin{equation}
	\begin{aligned}
		\lim\limits_{N\to+\infty} \mathbb{E} \mathsf{D}(1) &= \frac{1}{2}\int_{q_{+}(y)\textless q_{-}(y)}\Psi\left(\left\lvert \text{ln}\frac{q_{+}(y)}{q_{-}(y)}\right\rvert\right) q_{+}(y)dy\\
		&\quad +\frac{1}{2}\int_{q_{+}(y)\textgreater q_{-}(y)}\Psi\left(\left\lvert \text{ln}\frac{q_{+}(y)}{q_{-}(y)}\right\rvert\right)q_{-}(y)dy.  
	\end{aligned}     
\end{equation}
\subsection{Proof of Lemma 2}
Defining $\mathsf{W}_n = \frac{\mathsf{R}_n}{N}\mathbf{1}(\text{sgn}(\mathsf{T}_n )\cdot \mathsf{X}_n(1)\textless 0)$ and $\Tilde{\mathsf{W}}_n = \mathsf{W}_n - \mathbb{E}\mathsf{W}_n$, we have
\begin{equation}
	\label{varD1}
	\text{var}\mathsf{D}(1) = \frac{1}{N^2}\sum_{i = 1}^N\sum_{j = 1}^N \mathbb{E}[\Tilde{\mathsf{W}}_i\Tilde{\mathsf{W}}_j].
\end{equation}\par
We disassemble (\ref{varD1}) into two situations: $j=i$ and $j \neq i$, and then separately treat them following similar techniques as in \cite[Appendix C]{liu2022orbgrand}. The analysis reveals that $\text{var}\mathsf{D}(1)$ asymptotically vanishes as $N \rightarrow \infty$. 
\subsection{Proof of Lemma 3}
Since $\underline{\mathsf{T}}$ is induced by $\underline{\mathsf{X}}(1)$, it is independent of $\underline{\mathsf{X}}(m^{\prime})$, and we have
\begin{equation}
	\begin{aligned}
		\label{D(m)}
		& \mathbb{E}\left\{e^{N\theta \mathsf{D}(m^{\prime})} \Big\lvert \underline{\mathsf{T}}\right\}
		=\prod_{n = 1}^N \mathbb{E}\left\{ e^{\theta \frac{\mathsf{R}_n}{N}\mathbf{1}(\text{sgn}(\mathsf{T}_n)\cdot \mathsf{X}_n(m^{\prime})\textless 0)}\Big\lvert \underline{\mathsf{T}}\right\}.
	\end{aligned}
\end{equation}\par 
We can use similar approach as in \cite[Appendix D]{liu2022orbgrand}, exploiting the fact that $\mathsf{R}_n$ is determinisitc once $\underline{\mathsf{T}}$ is given, to obtain
\begin{equation}
	\begin{aligned}
		&\mathbb{E}\left\{ e^{\theta \frac{\mathsf{R}_n}{N}\mathbf{1}(\text{sgn}(\mathsf{T}_n)\cdot \mathsf{X}_n(m^{\prime})\textless 0)}\Big\lvert \underline{\mathsf{T}}\right\}=\frac{1}{2}(1 + e^{\theta \frac{\mathsf{R}_n}{N}}).
		\label{R}
	\end{aligned}
\end{equation}\par
Substituting (\ref{R}) into (\ref{D(m)}) and using the fact that $\left\{\mathsf{R}_n\right\}_{n=1,2,\cdots,N}$ is a permutation of $\{1, 2, \ldots, N\}$, we obtain
\begin{equation}
	\begin{aligned}
		\Delta(\theta)  = \int_{0}^{1}\text{ln}(1+ e^{\theta t})dt - \text{ln}2.
	\end{aligned}
\end{equation}
\section{}\label{B}
\section*{GMI of SGRAND}
The GMI of SGRAND can be calculated by the formula of GMI \cite[Eqn. (12)]{ganti2000mismatched} as 
\begin{equation}
	\begin{aligned}
		I_{\text{SGRAND}}=\sup\limits_{\theta \textless 0}\left\{\theta \mathbb{E}d(\mathsf{X},\mathsf{Y})-\mathbb{E}\left[\text{ln}\sum_{x \in \left\{+1,-1\right\}}\frac{e^{\theta d(x,\mathsf{Y})}}{2}\right] \right\},
	\end{aligned}
\end{equation} where $d(x,y)=\left\lvert \text{ln}\frac{q_{+}(y)}{q_{-}(y)}\right\rvert\cdot \mathbf{1}\left(\text{sgn}\left(\text{ln}\frac{q_{+}(y)}{q_{-}(y)}\right)\cdot x\textless 0\right)$.\par 
With some calculations, we have
\begin{equation}
	\begin{aligned}
		I_{\text{SGRAND}}&=\text{ln}2-\inf\limits_{\theta \textless0}\Bigg\{\mathbb{E}\left[\text{ln}\left(1+e^{\theta \left\lvert\text{ln}\frac{q_{+}(\mathsf{Y})}{q_{-}(\mathsf{Y})}\right\rvert}\right)\right]\\
		&\quad -\theta \cdot \frac{1}{2}\int_{q_{+}(y)\textless q_{-}(y)}\left\lvert\text{ln}\frac{q_{+}(y)}{q_{-}(y)}\right\rvert q_{+}(y)dy\\
		&\quad -\theta \cdot\frac{1}{2}\int_{q_{+}(y)\textgreater q_{-}(y)}\left\lvert\text{ln}\frac{q_{+}(y)}{q_{-}(y)}\right\rvert q_{-}(y)dy \Bigg\}.
		\label{SGRAND rate1}
	\end{aligned}
\end{equation}
\par 
\bibliographystyle{ieeetr}
\bibliography{ref.bib}
\end{document}